\documentclass{elsart}
\usepackage{graphicx}
\usepackage{graphicx}
\usepackage{epsfig}
\usepackage{amsmath}
\usepackage{amssymb}

\newcommand{\GeVcc}{\mbox{${\rm GeV/c^2}$}}

\newcommand{\hetrois}    {\mbox{$ ^{3}{\mathrm{He}}                            $}}
\newcommand{\xe}    {\mbox{$ ^{129}{\mathrm{Xe}}                            $}}
\newcommand{\ger}    {\mbox{$ ^{73}{\mathrm{Ge}}                            $}}
\newcommand{\al}    {\mbox{$ ^{27}{\mathrm{Al}}                            $}}
\newcommand{\fl}    {\mbox{$ ^{19}{\mathrm{F}}                            $}}

\newcommand{\neutt}{$\tilde{\chi}$}
\newcommand{\mchi}{${\rm M_{\tilde{\chi}}}$}
\newcommand{\mc}{${\rm M_{\chi^{\pm}}}$}
\newcommand{\omegachi}{$\Omega_{\tilde{\chi}}$}

\newcommand{\gams} {{$\gamma$-rays}}

\newcommand{\elec}{\mbox{${\rm e^{-}}$}}
\newcommand{\posi}{\mbox{$e^{+}$}}

\newcommand{\Mun}{\mbox{${\rm M_1}$}}
\newcommand{\Mde}{\mbox{${\rm M_2}$}}
\newcommand{\pbar}{\={p}}
\newcommand{\dbar}{\={D}}

\def\NIMA#1#2#3{{\rm Nucl.~Instr.~and~Meth.} {\bf{A#1}} (#2) #3}

\def\PLB{{\em Phys. Lett.}  B}

\def\PRD#1#2#3{{\rm Phys. Rev.} {\bf{D#1}} (#2) #3}

\def\JHEP#1#2#3{{\rm JHEP} {\bf{#1}} (#2) #3}

\def\PRL#1#2#3{{\rm Phys.~Rev.~Lett.} {\bf{#1}} (#2) #3}
\def\PLB#1#2#3{{\rm Phys.~Lett.} {\bf{B#1}} (#2) #3}
\def\APP#1#2#3{{\rm Astropart.~Phys.} {\bf{B#1}} (#2) #3}
\def\APJ#1#2#3{{\rm Astrophys.~J.} {\bf{#1}} (#2) #3}
\def\APJS#1#2#3{{\rm Astrophys.~J.~Suppl.} {\bf{#1}} (#2) #3}
\def\AA#1#2#3{{\rm Astron. \& Astrophys.} {\bf{#1}} (#2) #3}
\def\JCAP#1#2#3{{\rm JCAP} {\bf{#1}} (#2) #3}
\begin{document}
\begin{frontmatter}
\title{Supersymmetric dark matter search via spin-dependent interaction with \hetrois.}
\author{E. Moulin\thanksref{corr}},
\author{F. Mayet},
\author{D. Santos}
\thanks[corr]{Corresponding author : Emmanuel.Moulin@lpsc.in2p3.fr (phone: +33 4 76 28 41 49, fax:+33 4 76 28 00 04)}
\address{Laboratoire de Physique Subatomique et de Cosmologie, 
 CNRS/IN2P3 et Universit\'e Joseph Fourier, 
 53, avenue des Martyrs, 38026 Grenoble cedex, France}
\begin{abstract}
The potentialities of MIMAC-He3, a MIcro-tpc MAtrix of Chambers of Helium 3, 
for supersymmetric dark matter search 
are discussed within the framework of effective MSSM models
without gaugino mass unification at the GUT scale. A phenomenological study 
has been done to investigate the sensitivity of the MIMAC-He3 detector to 
neutralinos (\mchi \ ${\rm\gtrsim\,6\,\GeVcc)}$  via spin-dependent interaction with \hetrois~as 
well as its complementarity 
to direct and indirect detection experiments. 
Comparison with other direct dark matter searches will be presented 
in a WIMP model-independent framework.
\end{abstract}
\end{frontmatter}
\newpage
\section{\label{sec:intro}Introduction}
Since recent high accuracy experimental results in observational cosmology,
the existence of non-baryonic dark matter seems to be well established.
CMB results~\cite{wmap,archeops} used in combination with high redshift supernova~\cite{snap} 
and large scale structure surveys~\cite{sdss},
seem to point out that most of the matter in the Universe
consists of cold non-baryonic dark matter (CDM).
As a consequence, the range for CDM density has reached an unprecedented level of accuracy: 
${\rm \Omega_{CDM}\,h^2_0\,=\,0.12\,\pm\,0.04\,}$~\cite{wmap}, with ${\rm h_0\,=\,0.73\,\pm\,0.03}$ 
the normalized Hubble expansion rate~\cite{pdg}.
This non-baryonic cold dark matter consists of still not detected particles whose well motivated 
candidates are the WIMPs (Weakly Interacting Massive Particles).
Supersymmetric theories with R-parity conservation provide a suitable candidate, 
the lightest supersymmetric particle (LSP), which can significantly account for CDM~\cite{jungman}. 
In various SUSY scenarii, this neutral and colorless particle is the lightest 
neutralino \neutt.\\
Tremendous experimental efforts on a host of techniques have been made in the field of direct search of
non-baryonic dark matter~\cite{edelweiss,cdms,seidel,electronsmache3}.
Several detectors reached sufficient sensitivity to begin to test regions of the SUSY parameter space.
However, they are still limited by neutron interactions in the sensitive medium. 
Energy threshold effect combined with the use of a heavy target nucleus 
leads to significant sensitivity loss for light WIMPs.\\
The main purpose of this letter is to show the interest to perform an experimental effort
as MIMAC-He3 (MIcro-tpc MAtrix of Chambers of Helium 3)~\cite{idm2004,mimache3} in order to search 
for non-baryonic dark matter particles, within the framework  of effective MSSM  models 
without gaugino unification at GUT scale. A full estimation of the neutralino \neutt~detection rate in MIMAC-He3 has been performed including the computation 
of the spin-dependent cross-section on \hetrois. An emphasis on light neutralino sensitivity of MIMAC-He3 will be laid
as well as its complementarity with scalar and indirect detections. A comparison to other spin-dependent dark matter 
experiments will be shown in a WIMP model-independent framework.    
\section{\label{sec:MIMAC-He3}\hetrois~as sensitive medium for dark matter search}
As reported elsewhere~\cite{dm2000,nima,4micc}, the use of
\hetrois~is motivated by its privileged features for dark matter search compared with other 
target nuclei.
With \hetrois~being a spin 1/2 nucleus, a detector made of such a material will be sensitive to the spin-dependent
interaction, leading to a natural complementarity to existing detectors mainly sensitive to 
the spin-independent interaction.
For massive WIMPs, the maximum recoil energy depends very weakly on the
WIMP mass as the \hetrois~nucleus is much lighter (${\rm m({\hetrois})=2.81\,GeV/c^2}$). 
Therefore, the energy range in which all the sought events 
fall is ${\rm \lesssim 6 \,keV}$. 
Thus, the recoil energy 
range needs to be studied from energy threshold up to ${\rm 6 \,keV}$. This narrow range, for the searched 
events, represents a key point to discriminate
these rare events from the background.  
The \hetrois~presents in addition the following advantages with respect to other sensitive materials
for WIMPs detection:
\begin{itemize}
\item a very low Compton cross-section to gamma rays, two orders of magnitude weaker than in Ge : ${\rm 9\times10^{-1}}$ barns
for 10 keV \gams~compared to ${\rm 2\times10^{2}}$ barns in Ge, 
\item no intrinsic X-rays,
\item the neutron signature. The capture process ${\rm 
n\,+\,\hetrois\,\rightarrow\,p\,+^3H\,+\,764\,keV}$ 
gives very different signals with respect to the WIMP events.
\end{itemize}
This property is a key point for dark matter search as neutrons in underground laboratory are considered as the
ultimate background. Careful simulations and measurements of the neutron production induced by 
high energy cosmic muon interaction in the shielding are compulsory \cite{mimache3,kudryavtsev}.
Despite this, any dark matter detector should be able to separate a \neutt~event from the neutron background. 
Using energy measurement and electron-recoil discrimination,
MIMAC-He3 presents a high rejection for neutrons due to capture and multi-scattering of neutrons~\cite{mimache3}.
A detailled description of the MIMAC-He3 project can be found in a forthcoming paper \cite{mimache3}.
\section{\label{sec:theo}Theoretical framework} 
The potentiality of such a detector has been investigated in the framework of effective 
MSSM (Minimal SuperSymmetric Models) with no gaugino mass unification at GUT scale, thus extending LEP-allowed \neutt \ mass range. 
This work follows early SUSY calculations in 
\cite{mache3plb}, done with a model-dependent analysis within restrictive MSSM.
\subsection{\label{sec:eff}Effective MSSM with no gaugino mass unification} 
The lightest neutralino~\neutt~is a well motivated candidate as it corresponds to the LSP 
in many SUSY models~\cite{jungman}.
It is defined as the lowest mass linear combination of the supersymmetric partners of the U(1) and SU(2) gauge bosons, 
the bino ${\rm \tilde{B}}$, the wino ${\rm \tilde{W}}$, and the two Higgsinos ${\rm \tilde{H}^0_1, \tilde{H}^0_2}$ :
\begin{equation}
\rm{\tilde{\chi}\,=\,a\,\tilde{B}\,+\,b\,\tilde{W}\,+\,c\,\tilde{H}^0_1\,+\,d\,\tilde{H}^0_2}
\end{equation}
A standard assumption of supersymmetric models is the unification condition of the three gaugino masses  ${\rm M_i\,(i=1,2,3)}$
at the GUT scale (${\rm \sim\,2\times10^{16}\,GeV}$). 
This hypothesis implies at the electroweak scale (EW) the following relation between ${\rm M_1\,and\,M_2}$ :
${\rm M_1\,=\,5/3\,tan^2\,\theta_W\,M_2}$,
yielding the standard formula ${\rm M_1 \simeq 0.5 \, M_2}$ at the EW scale.
Under the unification condition, 
the lower bound on the lightest neutralino mass is found to be \mchi \ ${\rm \ge\,36\,GeV/c^2}$ derived from 
LEP2 analysis~\cite{pdg}. 
In order to explore the probable existence of lighter neutralinos,
supersymmetric models where ${\rm M_1\,and\,M_2}$ are considered as independent parameters have been
investigated~\cite{ellis}.
The unification constraint can be relaxed by the introduction of a free parameter R defined by : 
\begin{equation}
{\rm M_1\,\equiv\,R\,M_2}
\end{equation} 
Thus, the departure from the gaugino mass universality can be studied 
with ${\rm R\,<\,0.5}$~\cite{hooper,bottino,belanger}.
The neutralino \neutt \ can be lighter 
than the limit obtained in scenarii with gaugino mass unification~\cite{bottino}.
In non-universal cases,  the lower limit on the neutralino mass exists whatever the R value
but diminishes when $\rm R < 0.5$ becoming roughly
\mchi $\rm \ge R \times M_{\chi^{\pm}}$. 
The lightest neutralinos are typically found in SUSY models associated with \Mun$\ll$\Mde,$\mu$, where 
a significant higgsino fraction is required to obtain
relatively high cross-sections.  
\subsection{SUSY parameter space}
\label{subsec:par}
A minimal set of parameters has been used in the framework of an effective MSSM with no gaugino mass unification at GUT
scale. The supersymmetric parameter space consists of the following independent parameters : 
\begin{center}
${\rm M_2,\,\mu\,,M_0\,,M_A\,,tan\,\beta\,,A_{t,b}\,,R}$
\end{center}
with ${\rm M_2}$ the gaugino mass parameter,
${\rm \mu}$ the higgsino mass parameter, ${\rm M_0}$ the common scalar mass,
${\rm M_A}$ the pseudo-scalar Higgs mass, 
${\rm tan\,\beta}$ the Higgs VEV ratio, ${\rm A_{t,b}}$ the trilinear couplings and R the ratio characterizing the 
amount of non universality.
A large scan of the parameter space has been done with the DarkSUSY code~\cite{ds} in which the departure from 
the universality has been
implemented. This code is a numerical package for supersymmetric dark matter calculations including all resonances, 
thresholds and coannihilations for cross-sections. 
The ranges used for the scan on each parameter defined at electroweak (EW) scale are given in table~\ref{table1}.
The resulting scan 
corresponds to a total number of $\sim 25\,\times\,10^6$ models.
It also includes the usual scheme with U(1) and SU(2) gaugino mass unification at GUT scale, for which ${\rm R = 0.5}$
at the EW scale. 
\begin{center}
\begin{table}[ht]
\begin{center}
\begin{tabular}{cccc}
\\
\hline
\hline
Parameter&Minimum& Maximum&Number of steps\\
\hline
\hline 
M$_0$ (GeV)&100&1000&11\\
M$_2$ (GeV)&50&1000&20\\
M$_A$ (GeV)&100&1000&10\\
$\mid \!\mu\! \mid$ (GeV)&50&1000&20\\
\multicolumn{4}{c}{\raisebox{0pt}[12pt][6pt]{sign($\mu$)\,=\,$\pm$\,1}}\\
tan$\beta$&5&60&12\\
R&0.01&0.5&10\\
\multicolumn{4}{c}{\raisebox{0pt}[12pt][6pt]{A$_{t,b}$=0}}\\
\hline
\hline
\end{tabular}
\caption{\label{table1}Scan of the SUSY parameter space including the departure from U(1)
and SU(2) gaugino unifcation at GUT scale.
The total number is $\sim$\,25$\times10^6$ models.}
\end{center}
\end{table}
\end{center}
\subsection{Experimental constraints}
\label{sec:exp}
\subsubsection{\label{subsec:acc}Accelerator limits}
Standard bounds from colliders usually come from derivations in mSUGRA type model analysis~\cite{pdg}. In the non universal 
scheme, such constraints have to be redefined. All the limits given below are extracted from Ref.~\cite{pdg}.
LEP and Tevatron set mass bounds on supersymmetric charged particles giving the lower limit for chargino masses :   
\mc${\rm \,\gtrsim\,103\,GeV/c^2}$. 
In models where the pseudoscalar mass is heavy, the limit obtained from LEP2 on the lightest CP even Higgs mass
is ${\rm M_h\,\ge\,114\,GeV/c^2}$. Nevertheless, in models with a light pseudoscalar (${\rm M_A\,\le\,200\,GeV/c^2}$), 
the LEP2 constraint is relaxed and the absolute bound is ${\rm M_h\,\ge\,91.6\,GeV/c^2}$.
These contraints have been imposed to the SUSY models used for this study. 
For the \neutt \ mass, the commonly quoted and employed bound,  ${\rm 36\,GeV/c^2}$, is derived from the lower bound of the
chargino mass \mc \ determined at LEP2 under the assumption that the U(1) and SU(2)
gaugino masses ${\rm M_1\,and\,M_2}$ satisfy the standard relationship ${\rm M_1\,\simeq\,0.5\,M_2}$ 
at the EW scale. In such scenarii, it is only through the relation between the gaugino parameters \Mun~and 
\Mde~that the neutralino mass limit is obtained. Indeed, the lower bound on \mc \ converts into bounds on \Mde \ and $\mu$.
The relationship between \Mun \ and \Mde \ implies a lower bound on  \Mun, leading 
to the commonly used limit on \mchi.
In non universal case, an absolute lower limit on the \neutt \ mass cannot be obtained because 
the chargino and neutralino masses are uncorrelated and the lower limit on the neutralino mass diminishes with ${\rm R\,<\,0.5}$.
However, a  lower limit can be derived from the cosmological bound on the amount of 
CDM, the latter being roughly inversely proportional to the square root of \mchi~\cite{bottino}. 
\subsubsection{\label{subsec:ind}Indirect limits}
In order to take into account latest results of observational cosmology~\cite{pdg}, 
we required the SUSY models to yield a \neutt~relic density 
\omegachi \ with a dominant contribution to dark matter. 
Only models giving a \omegachi~in the following range are considered :  
\begin{center}
${\rm 0.02\,\le\,\Omega_\chi\,h^2_0\,\le\,0.15}$
\end{center}
The relic density is constrained in such a range in order to account for various effects.
First, it has been shown~\cite{belanger2}, that the uncertainty,  for all SUSY codes, 
in the relic density calculation can be quite large.
Indeed, mass differences of about 1\% in the input SUSY mass spectrum, lead to a spread in the
calculated densities of 10\%. The discrepancies can be even larger for some
SUSY parameter space region.
Concerning the lower bound, it should be highlighted that one wants to keep models which 
provide a non negligeable relic density, that is to say
neutralinos which can contribute significatively to the amount of  CDM, but which not 
necessarily fill the entire amount of CDM.
A lower bound on the \neutt~mass can be derived from the upper bound on the \neutt~relic density.
It is estimated to be ${\rm M_{\tilde{\chi}}\,\gtrsim\,6\,\GeVcc}$~\cite{bottino,belanger}.
Experimental constraints from accelerators on
the muon anomalous magnetic moment, ${\rm a_\mu\,\equiv\,(g_\mu-2)\,/\,2 }$,
and the ${\rm b\rightarrow s+\gamma}$ decay  are taken into account.
For the muon anomalous magnetic moment, we choose a limit taking into account
\elec\posi~and $\tau$ results~\cite{pdg} :
\begin{center}
${\rm -25\,\le\,a_{\mu}\,\times\,10^{10}\,\le\,69}$  
\end{center}
As many contributions have to be clarified, our range is conservative.
Concerning the branching ratio of the rare decay, ${\rm BR(b\,\rightarrow\,s\,+\,\gamma)}$, 
from CLEO, BELLE and ALEPH measurements~\cite{pdg}, models are required to give a prediction 
falling in the range :
 \begin{center}
${\rm 2.04\,\le\,BR(b\rightarrow s+\gamma)\,\times\,10^{4}\,\le\,4.42}$
\end{center}
\subsubsection{Astrophysical parameters}
\label{subsec:ast}
As far as the detection on Earth is concerned, a local halo model is required. 
A spherical isothermal distribution with standard galactic halo parameter have been used. The velocity distribution is
assumed to be a standard isotropic maxwellian distribution in the galactic frame
with an average rms velocity v truncated above the escape velocity of the galaxy. 
We assume no clumpy structures in the halo even if studies have been done in such cases~\cite{bergstrom}.
The set of astrophysical parameters,
commonly used~\cite{pdg}, for the SUSY parameter scan is :
\begin{center}
${\rm \rho_0\,=\,0.3\,GeV\,/c^2\,cm^{-3}}$ and
${\rm v\,=\,220\,km\,s^{-1}}$
\end{center}  
\section{Neutralino direct detection with \hetrois}
\subsection{Spin-dependent cross-section on \hetrois}
\begin{figure}[hp]
\begin{center}
\includegraphics[scale=1]{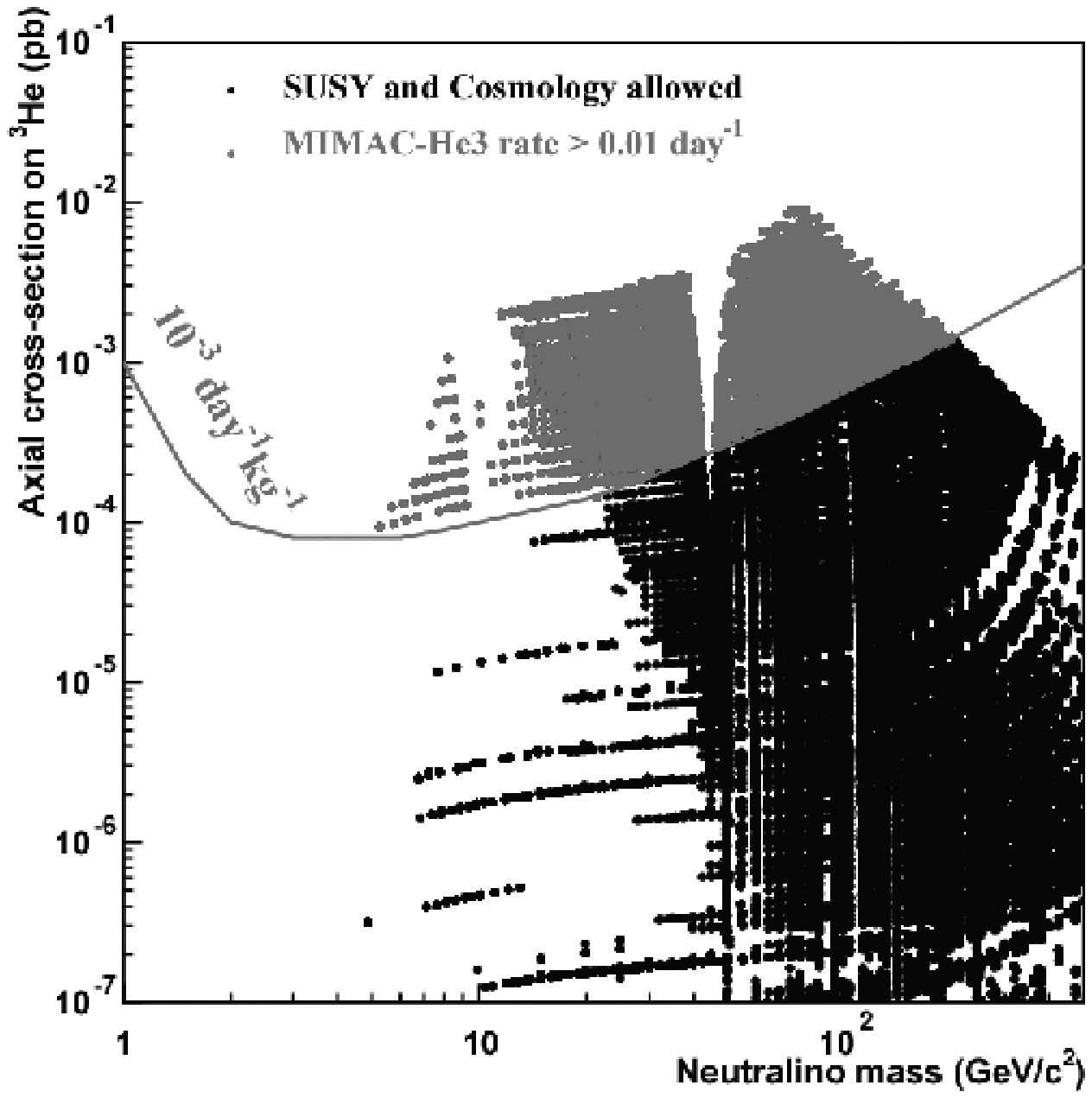}
\caption{\label{fig:SD}Spin-dependent cross-section on \hetrois~(pb) versus the neutralino mass (\GeVcc). 
Black points correspond to SUSY models allowed by collider and cosmological constraints (${\rm \sim 4\times10^5}$ models).
Projected exclusion curve for MIMAC-He3 (red solid line) with 10$^{-3}$ day$^{-1}$kg$^{-1}$background level is drawn. 
Models accessible by MIMAC-He3 correspond to red points (${\rm \sim 1.7\times10^5}$ models).}
\end{center}
\end{figure}

The neutralino \neutt~being a Majorana fermion, only the scalar (via H and h in t-channel and ${\rm \tilde{q}}$
in s-channel) and the axial (via Z in t-channel and ${\rm \tilde{q}}$ in s-channel)
interactions remain, the first one being called  spin-independent (SI) and 
the second one, the spin-dependent (SD) vanishing in the case of zero spin nuclei.
\hetrois~being an spin 1/2 nucleus is sensitive to the axial
interaction with the neutralino. Such an interaction is predominant on the 
scalar one for light nuclei such as \hetrois~\cite{jungman}.
The spin-dependent cross-section on \hetrois~has been evaluated in the whole
parameter space using the DarkSUSY code~\cite{ds}. The \neutt-quark 
elastic scattering amplitude are calculated via ${\rm \tilde{q}}$ or Z exchange. 
The 
amplitude on the nucleon is then evaluated by adding the contribution of each quark
weighted by the quark content of the nucleon. 
In order to obtain the cross-section on the nucleus, we will use the tree level expansion.
The expression at zero momentum transfer of the cross-section on \hetrois \ is then estimated as :
\begin{equation}
\rm{\sigma_{SD}(\hetrois)\,=\frac{32}{\pi}\,G_F^2\,\mu^2\,\frac{J+1}{J}(<S_p>a_p\,+<S_n>a_n)^2}
\end{equation}
where ${\rm G_F}$ is the Fermi coupling constant, ${\rm \mu}$ is the reduced mass of the \hetrois-\neutt~system, J the ground state angular momentum and
${\rm <S_{p,n}>}$ the proton and neutron spin content of \hetrois.
Using this expression, the \neutt~event rate has been calculated
for a 10 kg \hetrois~matrix 
and compared with the expected background 
rate. The result is presented on figure~\ref{fig:SD}
for SUSY models not excluded by collider constraints~(section~3.3.1) and providing a 
\neutt~relic density in the range of interest~(section~\ref{subsec:ind}) leading to ${\rm \sim 4\times10^5}$ allowed models
for our scan.
The projected exclusion curve for MIMAC-He3 corresponding to a
${\rm 10^{-3}\,day^{-1}kg^{-1}}$ background~\cite{mimache3} level is drawn. \\
The very low energy threshold (${\rm \sim 1 \ keV}$) \cite{mimache3} allows to be sensitive to light neutralinos \neutt~down to $\sim$ 6 \GeVcc~masses.
Models giving a neutralino rate higher than the expected background are
selected (above the exclusion curve), hereafter referred to as {\it accessible} to MIMAC-He3. 
They correspond to ${\rm \sim 1.7\times10^5}$ models.
It can be noticed that MIMAC-He3 would present a sensitivity for neutralino masses
from ${\rm\sim}$ 6 to 200 \GeVcc~for the expected background level.\\
When the \neutt~mass approaches 40~\GeVcc, the spin-dependent cross-section on \hetrois~suddenly decreases.
Indeed, the mass ${\rm M_Z/2}$  represents a resonance in the annihilation cross-section thus leading to a decrease in \omegachi.
The neutralinos associated to such models 
provide a relic density \omegachi~lower than 0.02. 
The maximum cross-section is ${\rm \sim 10^{-2}}$ pb  for \neutt~mass ${\rm\sim}$ 90~\GeVcc~and it decreases
with increasing masses.\\
\subsection{Complementarity with existing scalar detectors}
\begin{figure}[hp]
\begin{center}
\includegraphics[scale=1.]{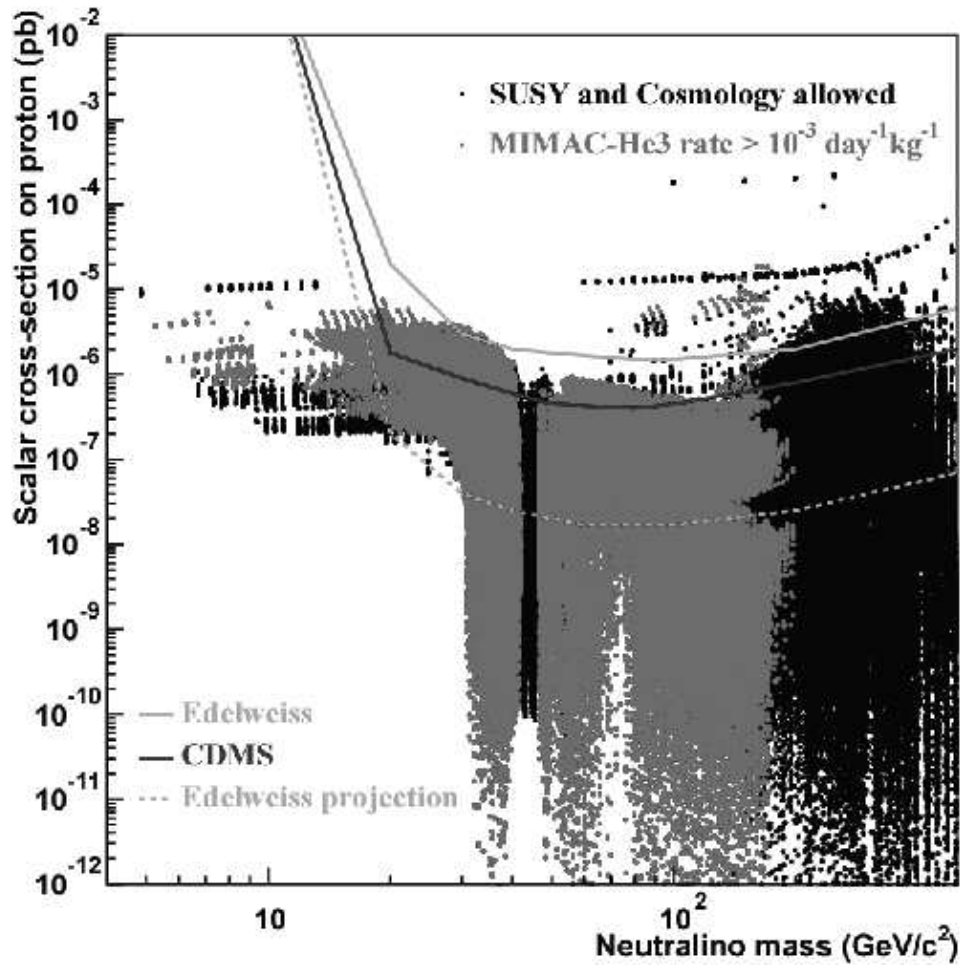}
\caption{\label{fig:SI}Scalar cross-section on proton (pb) versus the neutralino mass (\GeVcc). 
Black points region represent supersymmetric models satisfying both collider and cosmological constraints.
The models accessible by MIMAC-He3 correspond to red points. 
Exclusion curves from CDMS~\cite{cdms} and Edelweiss~\cite{edelweiss} experiments 
are plotted as well as the projected limit for Edelweiss (dotted line).}
\end{center}
\end{figure}

We study here the complementarity between the axial and the scalar interactions,
the latter one being widely exploited by most of the ongoing detectors. 
In order to compare the potentiality of MIMAC-He3 with respect 
to scalar detectors, the accessible models are 
distributed in the scalar cross-section on proton versus neutralino mass plane.  
Within the framework described in 
section~\ref{subsec:par}, the scalar cross-section on proton has been evaluated in the whole 
SUSY parameter space. The figure~\ref{fig:SI} presents the scalar cross-section on proton versus 
the \neutt~mass for SUSY models satisfying both collider and cosmological constraints.
An horizontal branch is observed towards light \neutt~masses corresponding to scalar cross-sections up to
${\rm 10^{-5}}$ pb. Models providing smaller cross-sections
imply too small values of \omegachi.\\ 
Exclusion curves from ongoing detectors have been plotted~\cite{edelweiss,cdms}
Models accessible to MIMAC-He3 have been projected in this diagram and it can be seen 
how they are distributed.
First, it can be noticed that SD and SI cross-section values can be extremely different.
Some SUSY models lie below the limits from scalar detectors 
(CDMS, Edelweiss) whereas they provide a \neutt~event rate in MIMAC-He3 higher than its 
background level. On the other hand, for such detectors, the energy threshold arises around 20~\GeVcc~leading to a 
significant loss in sensitivity for WIMP masses below. A large part of light \neutt, below 20~\GeVcc,
escapes from CDMS and Edelweiss detection whereas such a population is accessible to MIMAC-He3. 
Projected exclusion curves are also presented. Models visible by MIMAC-He3 remain unreachable
even for large mass detectors. This point highlights the complementarity between SD and SI 
direct dark matter detection.
\subsection{Complementarity with indirect detection}
\begin{figure}[hp]
\begin{center}
\includegraphics[scale=1.]{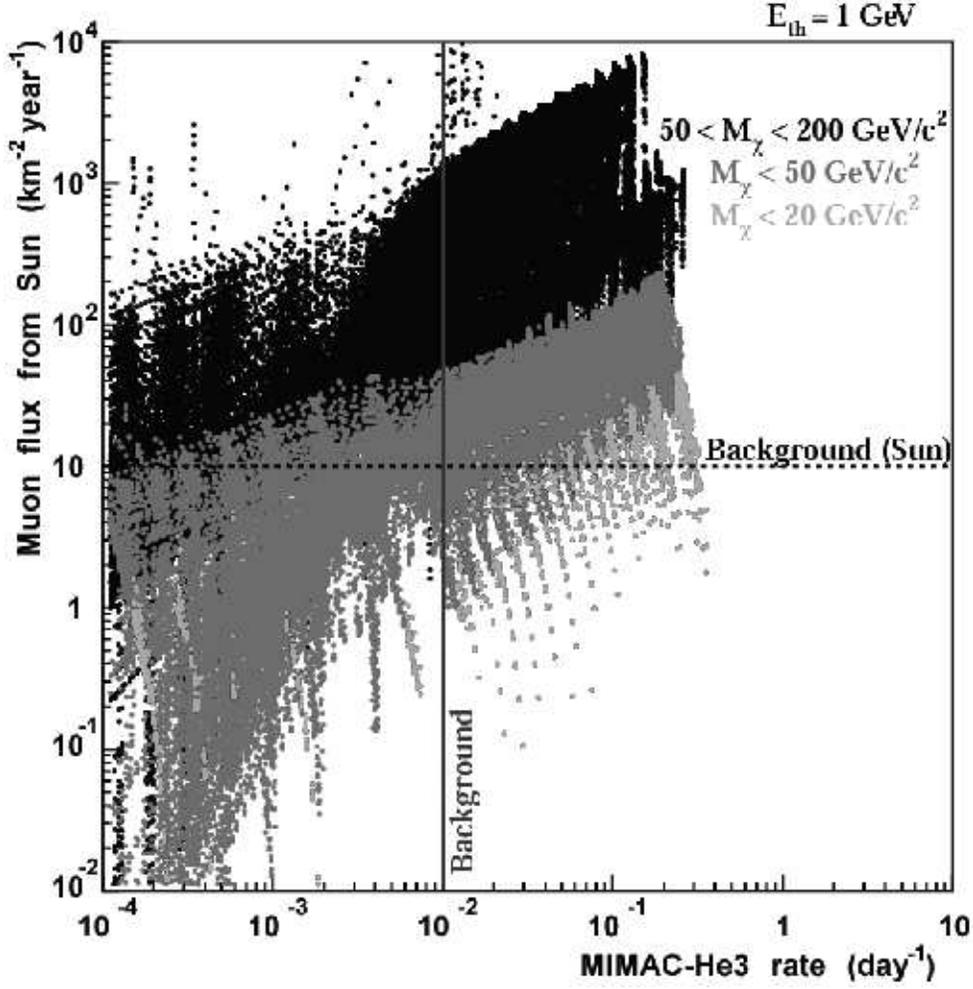}
\caption{\label{fig:compdirind}
WIMP-induced upward muon flux from Sun (${\rm km^{-2}\,year^{-1}}$) versus the \neutt~rate 
in MIMAC-He3 (${\rm day^{-1}}$) for different \neutt~mass ranges.
The muon energy threshold is 1 GeV. 
Background from energetic neutrinos produced in Sun's corona~\cite{seckel} (dashed line) and 
expected background  for MIMAC-He3 (solid lines) is indicated.}
\end{center}
\end{figure}

Indirect detection techniques are based on the measurement of particle flux (\posi, $\gamma$, \pbar, \dbar , $\nu$)
induced by pair annihilation of neutralinos. 
For instance, neutralinos can be captured gravitationally in celestial objects such as the Sun or the Earth and can annihilate 
in high energetic neutrinos~\cite{jungman}.
Earth signals are expected to be correlated to scalar direct detection due to the presence of even-even nuclei entering in Earth composition
whereas signals from the Sun being made of both odd and even nuclei (H, He) should be correlated to axial and 
scalar direct detections. 
In the theoretical framework described above (section 3.2),  
the possible complementarity of MIMAC-He3 with neutrino telescopes such as ANTARES~\cite{antares} or 
IceCube~\cite{icecube} is investigated.\\
The results are presented in terms of upward muon  fluxes above an energy threshold of 1 GeV~\cite{antares}
which will be the threshold of the next generation of the km$^2$-size neutrino telescopes.
Figure~\ref{fig:compdirind} shows the muon flux coming from the Sun (${\rm km^{-2}\,year^{-1}}$) versus the 
\neutt~rate in MIMAC-He3 (${\rm day^{-1}}$), for a 10 kg detector. 
The background for the neutrino telescopes is given by cosmic rays interacting 
in the Sun's corona. It is usually taken at ${\rm 10\,km^{-2}year^{-1}}$~\cite{seckel}. As a comparison, latest Super-Kamiokande results
 on WIMP-induced upward muon flux from Sun \cite{superk} (for a 50 \GeVcc \ WIMP) is of the order of 
$\rm 1.6 \times 10^{3} \ km^{-2}\,year^{-1}$, but with an energy threshold of 18 \GeVcc .
In the case of MIMAC-He3, the background level is considered at ${\rm 10^{-3}\,day^{-1}kg^{-1}}$. 
Four regions can be separated in this figure. 
First, some models lie below the 2 expected background levels : they are not accessible.
On the contrary, models in the upper right part of the figure should be seen via both detection methods.
Light neutralinos are expected to be visible only in MIMAC-He3 because they yield a muon flux
well below the background level. 
\subsection{Complementarity with spin-dependent direct detection experiments}
In the case of the SI interaction, the limits on the cross-section on the nucleus can be translated 
to bounds on the WIMP-proton cross-section. The conversion is relatively straightforward given 
the fact that the
WIMP coupling to the nucleus is proportional to the square of the nucleus mass.
The comparison of experimental results concerning the SD interaction is more problematic. 
Since the spin of the target nucleus is carried both by constituent protons and neutrons,
when converting to a WIMP-proton cross-section a value of the ratio of the 
WIMP-proton and WIMP-neutron cross-sections must be assumed. But this ratio can vary significantly 
depending on the WIMP composition.
Recently, a model-independent method~\cite{tovey} has been developed to enable 
comparison among SD direct searches of dark matter. 
The WIMP-nucleon cross-sections ${\rm \sigma_{p,n}^{lim(A)}}$ are given by~\cite{tovey}, in the limit ${\rm a_{n}=0}$ (resp. ${\rm a_{p}=0}$) :
\begin{equation}
\sigma_{p,n}^{lim(A)}\,=\,\frac{3}{4}\,\frac{J}{J+1}\,\frac{\mu^2_{p,n}}{\mu_A^{2}}\,\frac{\sigma_A}{<S_{p,n}>^2}
\end{equation}
where ${\rm \mu^2_{p,n}}$ is the WIMP-proton (resp. neutron) reduced mass and ${\rm \sigma_A}$ the WIMP-nucleus cross-section 
limit deduced from experiment.\\
As shown in~\cite{tovey}, the allowed values of ${\rm a_{p,n}}$, for a given WIMP mass, are required to fall in the inside region defined by :
\begin{equation}
a_p\,\le\,-\,\frac{<S_n>}{<S_p>}\,a_n\,\pm\,\sqrt{\frac{\pi}{24G_F^2\mu_p^2}\,\sigma_p^{lim(A)}}
\end{equation}
In such a framework, exclusion curves are presented in the (${\rm a_p,a_n}$) plane
for a given WIMP mass ${\rm M_{\tilde{\chi}}}$, where
${\rm a_p}$ and ${\rm a_n}$ are respectively the effective proton and neutron coupling strengths. 
Consistent exclusion curves are 3-dimensional plot : ${\rm a_p}$, ${\rm a_n}$ and ${\rm M_{\tilde{\chi}}}$.
For detectors with one active nucleus,
they are composed of two straight lines for which the gradient is given by the ratio ${\rm -\!<\!S_n\!>\!/\!<\!S_p\!>}$. Allowed
values of ${\rm a_p}$ and ${\rm a_n}$ lie in between these two straight lines. 
Consequently, the values of the spin contents of target nuclei are a key point for SD detection. Commonly used values are recalled in 
table~\ref{table2} for the reader's convenience.
\begin{center}
\begin{table}[ht]
\begin{center}
\begin{tabular}{ccc}
\hline
\hline
${\rm Nucleus}$&${\rm<S_p>}$&${\rm<S_n>}$\\
\hline
\hline
\hetrois&-0.050&\bfseries{0.490}\\
${\rm ^{19}F}$&\bfseries{0.477}&-0.004\\
${\rm ^{27}Al}$&\bfseries{0.343}&0.030\\
${\rm ^{73}Ge}$&0.030&\bfseries{0.378}\\
${\rm ^{129}Xe}$&0.028&\bfseries{0.359}\\
\hline
\hline
\end{tabular}
\caption{\label{table2}Proton and neutron spin content values for various nuclei. Values are taken from~\cite{tovey}
and the references therein. For discussions, see~\cite{bednyakov}.}
\end{center}
\end{table}
\end{center}

\begin{figure}[hp]
\begin{center}
\mbox{\hspace{-.5cm}\includegraphics[scale=0.4]{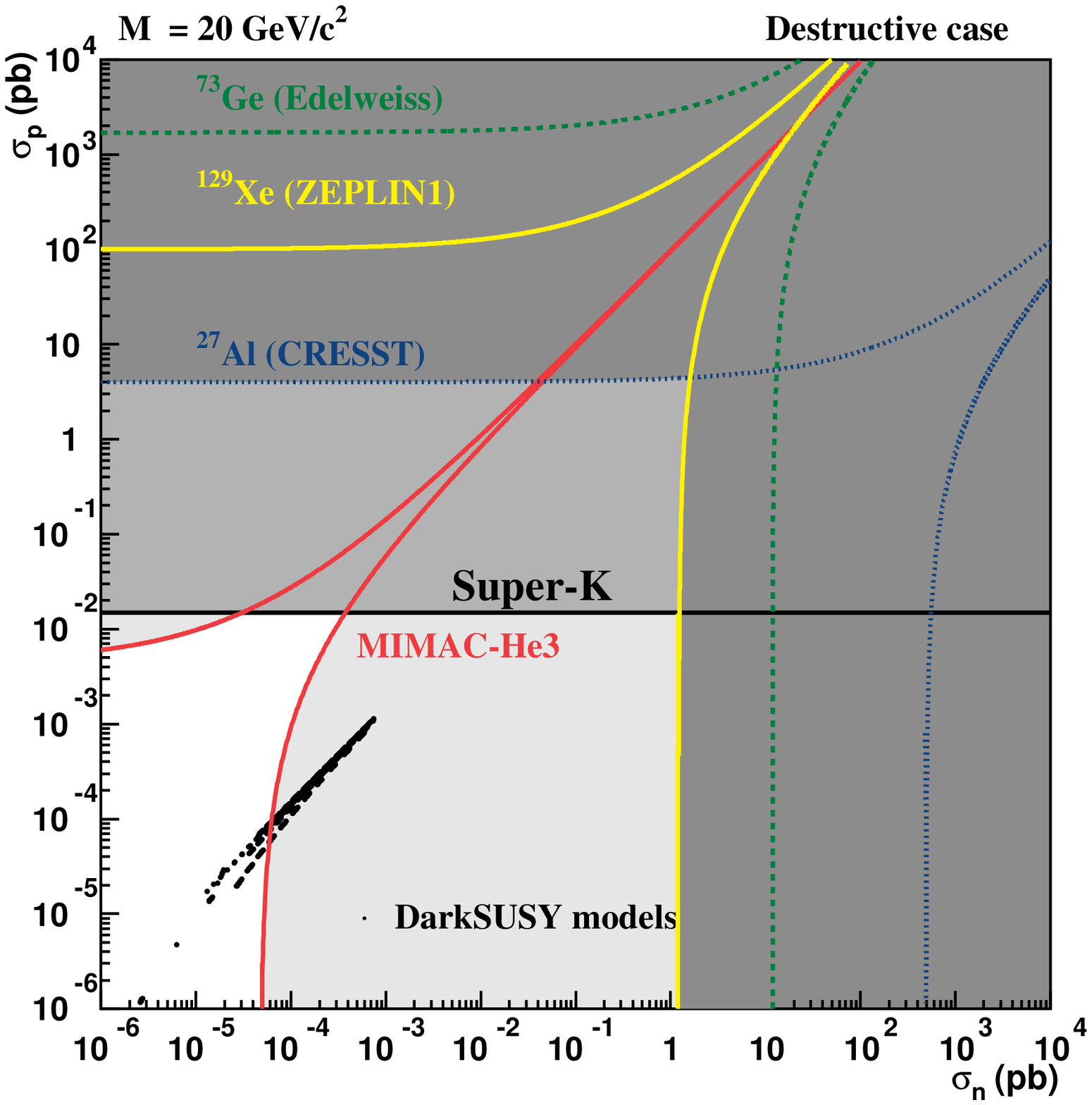}
\includegraphics[scale=0.4]{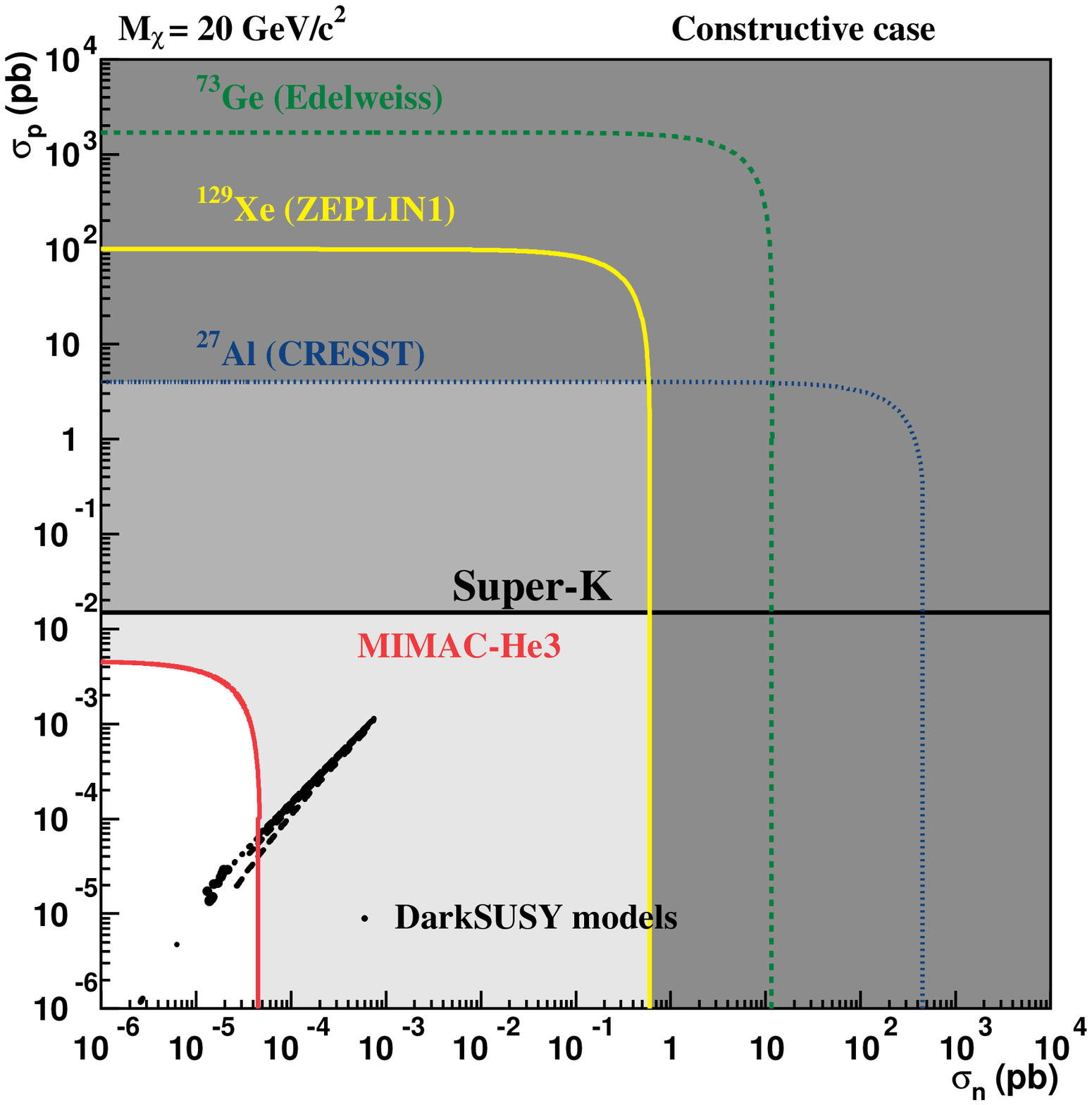}}
\mbox{\hspace{-.5cm}\includegraphics[scale=0.4]{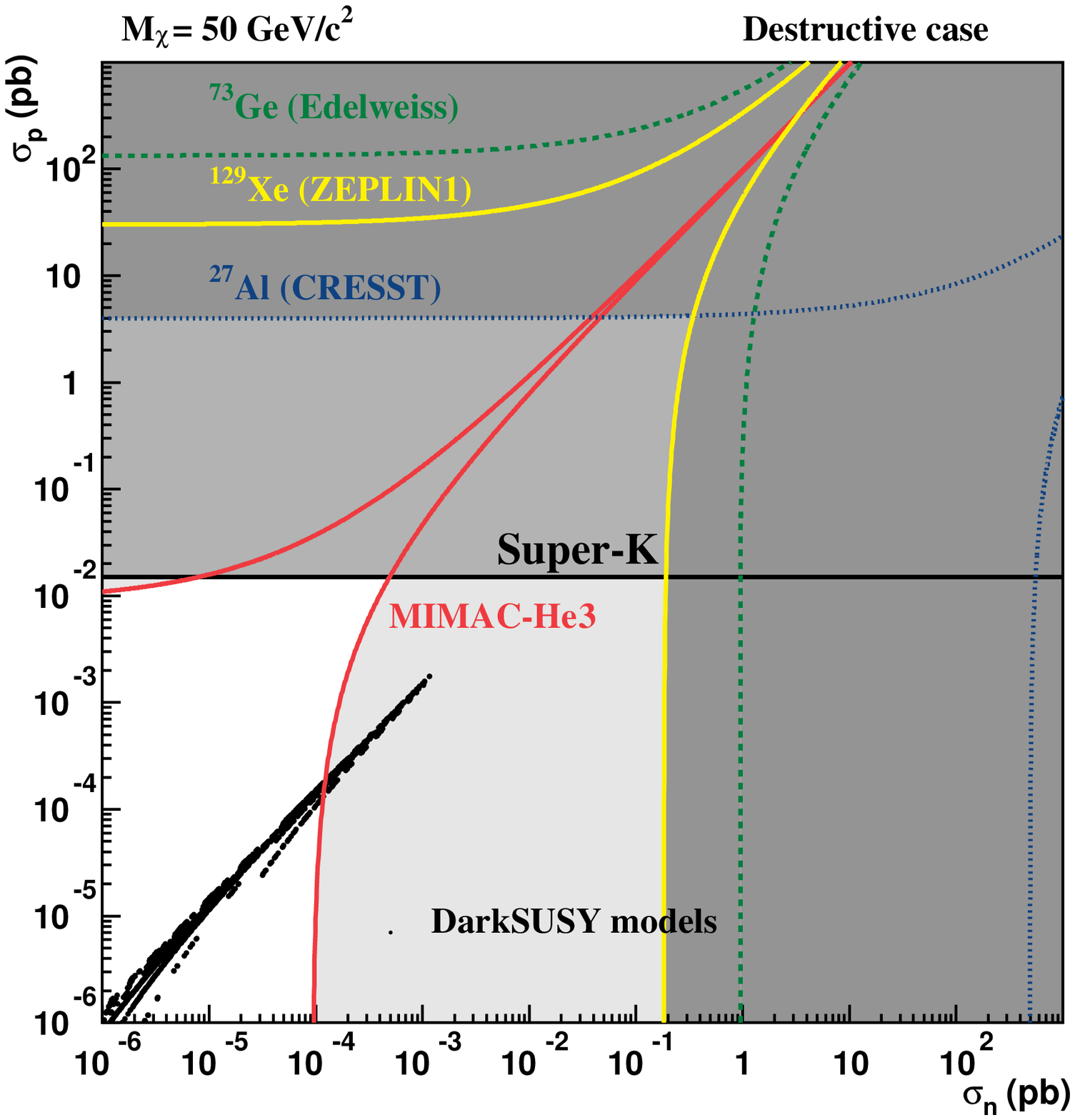}
\hspace{.25cm}\includegraphics[scale=0.4]{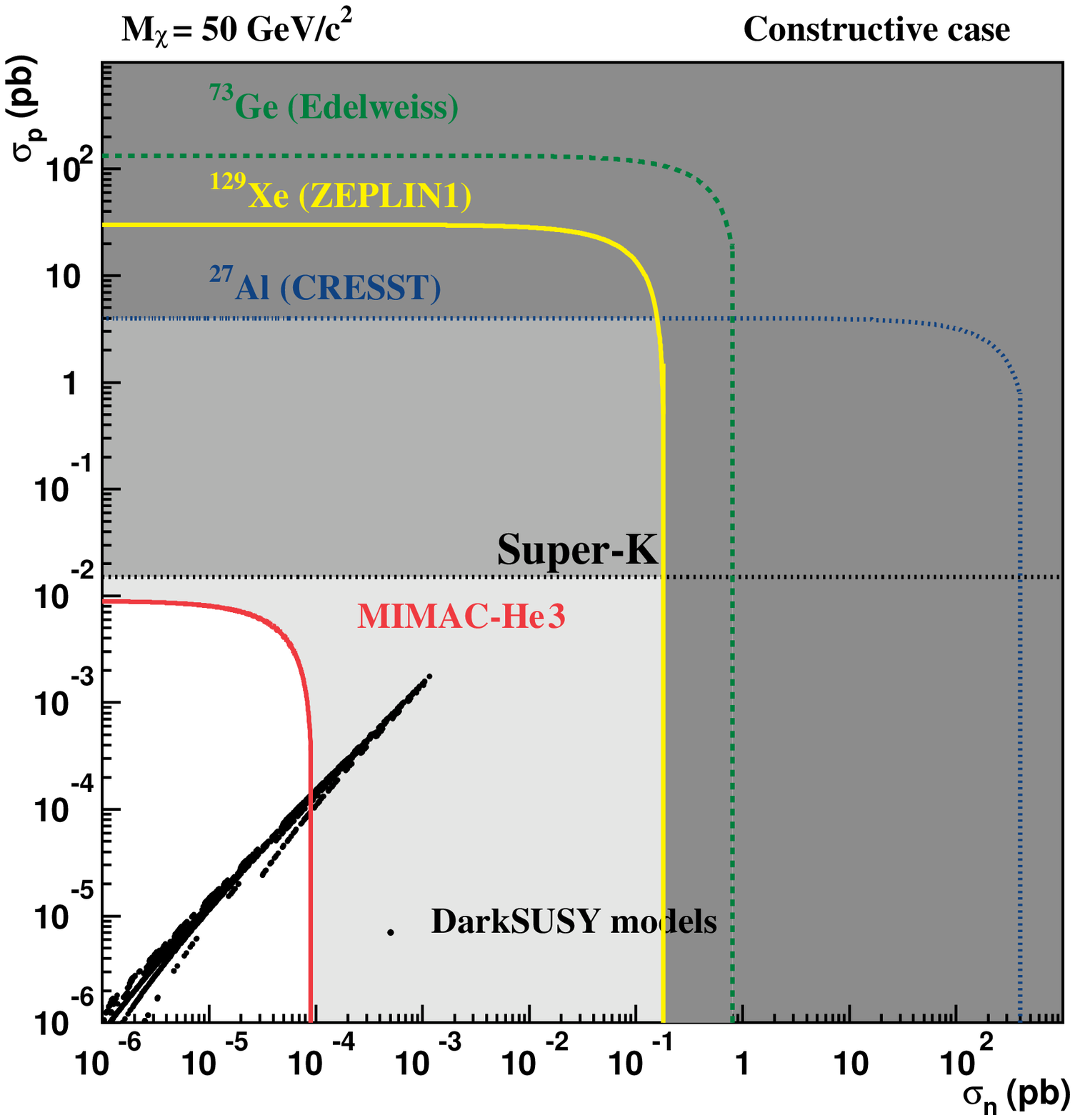}}
\caption{\label{fig:sigpsigndirind} SUSY models satisfying both accelerator and 
cosmological constraints (black points) are presented 
in the cross-section on proton (pb) versus cross-section on neutron (pb) diagram
for 20 \GeVcc \ (upper plots) and 50 \GeVcc \ (lower plots) neutralino masses
in the case of destructive interferences (left side plots) and constructive interferences  (right side plots). 
The projection exclusion curve of MIMAC-He3 corresponds to the red solid line. Exclusion curves 
from Edelweiss $^{73}$Ge (dashed line), CRESST $^{27}$Al (dotted line) and ZEPLIN-I \  \xe 
~\cite{seidel,edel,zeplin} are plotted.
The limit from Super-K \cite{superk} is also displayed (solid black line).
Dark gray corresponds to region currently excluded by SD direct detection, medium gray when adding the Super-K constraint 
and light gray with projected MIMAC-He3 exclusion.}
\end{center}
\end{figure}

Figure~\ref{fig:apan} shows exclusion curves for ongoing spin-dependent
searches in the (${\rm a_p,a_n}$) plane for selected 20 (left side) and 50 \GeVcc (right side) \ WIMP masses :  
CRESST-\al~\cite{seidel} (dashed line), Edelweiss-\ger~\cite{edel} (dash-dotted line), 
SIMPLE-\fl~\cite{collar}  (dotted line) and ZEPLIN-I-\xe~\cite{zeplin}  (yellow solid line).
In the upside plots,  the current excluded region (dark gray) is defined, for each WIMP mass,
by the intersection of the most constraining exclusion curves. 
It should be noticed that in the 20 \GeVcc \  mass case, 
Edelweiss constraint does not appear in this ${\rm (a_p,a_n)}$ range. 
The MIMAC-He3 projection (red solid line) allows to further constrain the current allowed region leading to a projected 
excluded region  defined by the light gray region. Only the white region should be left allowed with MIMAC-He3. 
The downside plots present a zoom for WIMP masses of 20 and 50~\GeVcc .
It can be seen that MIMAC-He3 should allow to put much stronger constraints on the 
SUSY region, especially for light WIMP masses for which other neutron based detectors provide much weaker constraints, 
due to threshold and target nucleus mass effects.\\
In addition to be a model-independent framework, the (${\rm a_p,a_n}$) diagram highlights the natural complementarity 
between  various SD detectors. In particular, proton based (${\rm
^{19}F,\,^{23}Na,\,^{27}Al,^{35}Cl,\,^{127}I}$) and 
neutron based detectors (\hetrois, ${\rm ^{73}Ge}$, ${\rm ^{129}Xe}$) provide orthogonal constraints. 
Furthermore, the sign of the ${\rm \!<\!S_n\!>\!/\!<\!S_p\!>}$ ratio governs the sign of the slope thus giving another complementarity 
among neutron based detectors. \\
However, due to the wide range for ${\rm a_p}$ and  $\rm a_n$, between experimental constraints and expected SUSY models, a more 
convenient representation would be obtained in the nucleon SD cross-section plane (${\rm \sigma_p, \sigma_n}$). Two cases 
should then be distinguished depending on the sign of the ratio  ${\rm <\!S_p\!>a_p/\!<\!S_n\!>a_n}$~:
the constructive (positive sign) or the destructive (negative sign) interferences.\\
Figure~\ref{fig:sigpsigndirind} shows the result in the (${\rm \sigma_p, \sigma_n}$) plane 
in the destructive (left side) and constructive (right side) interference cases, 
for 20 \GeVcc \ (upper side) and 50 \GeVcc \ (down side) neutralinos. 
For a given exclusion limit, the excluded region lies outside the two curves.
First, it should be noticed that, for a given experiment, there is a rather large difference between the 
two types of interferences in terms of experimental
constraints. Indeed, in the destructive case, some models with both high neutron and proton cross-sections may still remain
unreachable to a given target nucleus, i.e. models lying inside the "funnel".\\
Several exclusion curves are presented on figure~\ref{fig:sigpsigndirind}, such as CRESST (dotted line), ZEPLIN-I
(yellow solid line) and  
Edelweiss (dashed line). These curves are obtained from calculation based 
on~\cite{seidel,edel}.\\
The current excluded region (light gray) in this plane is given by the 
combination of theses curves. It also includes the limit from indirect DM detection ($\nu$ telescopes). 
As said previously, the capture rate in the Sun and thus the
neutrino flux is exclusively sensitive to the SD cross-section on proton. Therefore, a band 
along the ${\rm \sigma_n}$ axis is obtained. 
The Super-K limit  (black solid line) is displayed~\cite{superk} on figure~\ref{fig:sigpsigndirind}. The constraint from this 
experiment strongly reduces the allowed region. A near orthogonality is obtained with neutron based experiments.
On the other hand, proton based experiments (CRESST) are well overlapped by Super-K limit.\\
However, SUSY models (black points) neither excluded by accelerator  nor 
cosmological constraints lie well below this limit. All the models from a large 
SUSY parameter scan including non-universal models,  correspond to higgsino-like neutralinos for 
which ${\rm a_p/a_n}$ is of the order of 1.5. In the case of gaugino neutralinos,
this ratio can vary by several orders of magnitude as demonstrated in \cite{lewin}.\\
The projected exclusion curve for MIMAC-He3 is displayed, showing a strong constraint on the 
SD cross-section on neutron. It can be seen that most of 20 \GeVcc \ 
neutralinos, escaping from detection of ongoing experiments, would be visible by MIMAC-He3. 
A large part of 50 \GeVcc \ neutralinos would also be accessible.
In conclusion, MIMAC-He3 would present a sensitivity to SUSY models allowed by 
present cosmology and accelerator constraints. This study  highlights the complementarity of this experiment with 
most of current SD experiments : proton based detectors as well as $\nu$ telescopes. 
\section{Conclusion}
It has been shown that a 10 kg \hetrois~detector with a threshold of about 1 keV and with electron-recoil discrimination
(MIMAC-He3) would allow to reach
in a significant part of the SUSY parameter space, a \neutt~event rate higher than the estimated background.
MIMAC-He3 would be sensitive to SUSY models excluded neither by collider limits nor by neutralino
relic density constraint. This new project would be sensitive to SUSY regions not accessible by 
ongoing scalar detectors. 
The complementarity of MIMAC-He3 with ongoing
experiments is due to its sensitivity to spin-dependent interaction 
and its light target nucleus allowing to detect light neutralinos which are out of reach
of existing detectors.\\

\noindent \textbf{Acknowledgments : }\\
The authors wish to thank M. Bastero-Gil for fruitful discussions.


\newpage

\newpage

\newpage
 
\newpage
\begin{figure}[hp]
\begin{center}
\mbox{\hspace{-1cm}\includegraphics[scale=0.8]{apan20-50GeVnz.epsi}}
\mbox{\hspace{-1cm}\includegraphics[scale=0.8]{apan20-50GeVz.epsi}}
\caption{Exclusion curves for spin-dependent direct dark matter searches
in the (${\rm a_p, a_n}$) plane. 
The exclusion curves from CRESST-\al~\cite{seidel} (dashed line), Edelweiss-\ger~\cite{edel} (dash-dotted line), 
SIMPLE-\fl~\cite{collar}  (dotted line)  and ZEPLIN-I-\xe~\cite{zeplin} are plotted.
In the upside plots, the current excluded regions (dark gray) are defined 
by the  intersection of the most constraining exclusion curves (ZEPLIN-I and CRESST). 
The MIMAC-He3 projection (red solid line) allows to further constrain this region leading to a projected 
excluded region  defined by the light gray region. Only the white region should be left allowed with MIMAC-He3. 
The downside plots present a zoom for WIMP masses of 20 \GeVcc \ (left) and 50 \GeVcc \ (right).}
\label{fig:apan}
\end{center}
\end{figure}

\newpage

\end{document}